%% file: main.tex
 \documentclass[pmlr,twocolumn,10pt]{jmlr} 

\usepackage{booktabs}
\usepackage{siunitx}

\newcommand{\equal}[1]{{\hypersetup{linkcolor=black}\thanks{#1}}}

\theorembodyfont{\upshape}
\theoremheaderfont{\scshape}
\theorempostheader{:}
\theoremsep{\newline}

\jmlrworkshop{Machine Learning for Health (ML4H) 2023} 

 \title[Is Ignorance Bliss? The Role of Post Hoc Explanation Faithfulness and Alignment in Model Trust]{Is Ignorance Bliss? Investigating How Faithfulness and Alignment of Post Hoc Explanations Affect Model Trust in Laypeople and Domain Experts}

\author{%
\Name{Tessa Han}\equal{These authors contributed equally} \Email{than@g.harvard.edu}\\
\addr Harvard University, Cambridge, MA
\AND
\Name{Yasha Ektefaie}\footnotemark[1] \Email{yasha\_ektefaie@g.harvard.edu }\\
\addr Harvard University, Cambridge, MA
\AND
\Name{Maha Farhat} \Email{maha\_farhat@hms.harvard.edu}\\
\addr Harvard University, Cambridge, MA
\AND
\Name{Marinka Zitnik} \Email{marinka@hms.harvard.edu}\\
\addr Harvard University, Cambridge, MA
\AND
\Name{Himabindu Lakkaraju} \Email{hlakkaraju@hbs.edu}\\
\addr Harvard University, Cambridge, MA
}

\begin{document}

\maketitle

\begin{abstract}
Post hoc explanations have emerged as a way to improve user trust in machine learning models by providing insight into model decision-making. However, explanations tend to be evaluated based on their alignment with prior knowledge while the faithfulness of an explanation with respect to the model, a fundamental criterion, is often overlooked. Furthermore, the effect of explanation faithfulness and alignment on user trust and whether this effect differs among laypeople and domain experts is unclear. To investigate these questions, we conduct a user study with computer science students and doctors in three domain areas, controlling the laypeople and domain expert groups in each setting. The results indicate that laypeople base their trust in explanations on explanation faithfulness while domain experts base theirs on explanation alignment. To our knowledge, this work is the first to show that (1) different factors affect laypeople and domain experts' trust in post hoc explanations and (2) domain experts are subject to specific biases due to their expertise when interpreting post hoc explanations. By uncovering this phenomenon and exposing this cognitive bias, this work motivates the need to educate end users about how to properly interpret explanations and overcome their own cognitive biases, and motivates the development of simple and interpretable faithfulness metrics for end users. This research is particularly important and timely as post hoc explanations are increasingly being used in high-stakes, real-world settings such as medicine.
\end{abstract}

\input{1-intro}
\input{2-related-work}
\input{3-preliminaries}
\input{4-problem-statement-and-method}
\input{5-exp-results}

\input{6-conclusion}

\bibliography{bibliography.bib}
\onecolumn
\input{7-appendix}
\vspace{-10cm} 
\end{document}

%% file: 1-intro.tex
\section{Introduction}
\label{sec:intro}

Post hoc explanations have emerged as a way to improve user trust in machine learning models by providing insight into their decision-making \citep{arrieta2020explainable}, particularly in high-stakes settings such as medicine \citep{yu2018artificial}, law \citep{walters2021cyber}, and finance \citep{cao2022ai}. These explanations are often evaluated based on their alignment with human knowledge (“alignment”) \citep{smilkov2017smoothgrad, sundararajan2017axiomatic, saporta2022benchmarking}. The more explanations align with the knowledge of domain experts, the more accurate they are deemed to be. However, the faithfulness of an explanation with respect to the model (“faithfulness”), a fundamental criterion, is often overlooked. Explanation faithfulness is an important criterion when evaluating post hoc explanations because an explanation may seem valid to a user based on its alignment with domain knowledge, but completely misrepresent what happens in a model. Furthermore, the effect of explanation faithfulness and alignment in user trust and whether this effect differs among laypeople and domain experts is unclear.

Therefore, this study focuses on the following two research questions: (1) For a post hoc explanation, how do the faithfulness and the alignment of an explanation influence user trust in the model? (2) How does this phenomenon differ for laypeople versus domain experts?

To address these questions, we conduct a user study examining how post hoc explanations affect model trust for laypeople and domain experts. In doing so, we aim to understand how explanation faithfulness and alignment influence user trust and how the presence of highly-specialized domain knowledge influences how users reason about explanation faithfulness and alignment. This work is significant because, by understanding factors that influence user trust in models, how these factors differ among laypeople and domain experts, we can assess whether these users (who often are not machine learning experts) are basing their trust in correct factors. This is especially critical in high-stakes settings where domain experts may use post hoc model explanations to make consequential decisions.

We find that, when determining the extent to which to trust models using post hoc explanations, laypeople base their trust on explanation faithfulness while domain experts base their trust on explanation alignment. To our knowledge, this work is the first to show that (1) different aspects of post hoc explanations affect laypeople and domain experts' trust in a model and (2) domain experts are subject to specific biases due to their expertise when interpreting post hoc explanations. By uncovering this phenomenon and exposing this cognitive bias, this work motivates the need to educate end users about how to properly interpret explanations and overcome their own cognitive biases, and motivates the development of simple and interpretable faithfulness metrics for end users. This research is particularly important and timely as post hoc explanations are increasingly being used in high-stakes, real-world settings such as medicine.

%% file: 2-related-work.tex
\section{Related Work}
\label{sec:relwork}

This work is grounded in and builds upon prior work in the research areas below. For each area, we discuss relevant prior work, highlight research gaps, and explain how this work addresses these gaps.

\textit{Role of domain knowledge in AI trust.} Prior works have examined the role of domain knowledge on user trust in AI systems and model explanations, finding that it tends to decrease user trust \citep{ferreira2020people, wang2021explanations, bayer2021role, dikmen2022effects}. However, prior work has not studied the role of domain knowledge on user trust in models in a medical setting, nor has prior work investigated how faithfulness and alignment with prior knowledge affect user trust. Therefore, this study investigates the effect of explanation faithfulness and alignment on user trust in models, and how domain knowledge modulates this effect, by directly comparing the behavior of laypeople and medical domain experts.

\textit{Faithfulness of explanations.} Prior works have investigated the faithfulness of explanation methods with respect to the underlying model. For example, \cite{adebayo2018sanity} found that gradient-based explanations are not always faithful to the underlying model. Other works have developed approaches to measure explanation faithfulness, such as through the remove and re-train method \citep{hooker2019benchmark} and the local function approximation approach \citep{ribeiro2016should, han2022explanation}. However, to our knowledge, explanation faithfulness has not been studied in the context of user trust. Therefore, this study investigates the effect of explanation faithfulness on user trust and how this effect may differ between laypeople and domain experts.

\textit{Alignment of explanations with prior knowledge.} Prior works have used the alignment of model explanations with prior knowledge to evaluate explanation quality. For example, SmoothGrad \citep{smilkov2017smoothgrad} and Integrated Gradients \citep{sundararajan2017axiomatic}, two popular explanation methods, qualitatively demonstrate that their saliency maps are visually cleaner than those generated by other methods. In the medical setting, prior work evaluated model explanations by comparing top features obtained from explanations with those determined by human experts \citep{saporta2022benchmarking, duell2021comparison}. However, these assessments of the human-alignment of explanations were performed without first examining explanation faithfulness, a foremost criterion of explanation quality. Furthermore, the effect and interplay of faithfulness and alignment with prior knowledge on user trust has not been previously studied. Therefore, this study investigates the roles of faithfulness and alignment with prior knowledge in establishing user trust in models and how domain knowledge may alter these roles.

%% file: 3-preliminaries.tex
\section{Preliminaries}

In this section, we explain concepts and terminology relevant to the study.

\textit{Explanation faithfulness.} The faithfulness of an explanation with respect to the model is the degree to which the explanation accurately reflects the behavior of the model. This paper focuses on post hoc explanations that are feature attributions. An explanation is faithful if its top features are indeed features that are important to the model's prediction and its bottom features are indeed less important to the model's prediction, and the explanation is unfaithful if these conditions are not met. 

While prior works have investigated explanation faithfulness (as discussed in Section~\ref{sec:relwork}), to our knowledge, there does not exist an interpretable, single-value metric to measure faithfulness. Thus, we develop such a faithfulness metric that is grounded in prior work. Specifically, under the local function approximation framework \citep{han2022explanation} explanations are interpretable models that perform local function approximation of the black-box model over a perturbation neighborhood around a given data point. Using this framework, we develop a faithfulness metric, called the F-score, where explanation faithfulness is the proportion of points in the perturbation neighborhood of an explanation for which the interpretable model has the same or similar output as the black-box model (where similar indicates that the two outputs differ by less than a user-specified amount). Thus, the F-score ranges between 0 and 1, with higher values indicating higher explanation faithfulness.

\textit{Explanation alignment.} The alignment of an explanation with respect to prior knowledge is the extent to which salient features indicated by the explanation are consistent with prior knowledge. For example, to explain the prediction of a dog detector model for an image containing a dog, an explanation is aligned if it highlights image regions corresponding to the dog and is unaligned if it highlights image regions irrelevant to the dog (e.g., the background).

%% file: 4-problem-statement-and-method.tex
\section{Research Questions}

This study investigates the following two research questions: (1) For a post hoc explanation, how do the faithfulness and the alignment of an explanation influence user trust in the explanation? (2) How does this phenomenon differ for laypeople versus domain experts?

\section{Methodology}

To investigate the research questions above, we conduct a user study with two participant groups, computer science students and doctors, in three domain knowledge settings. In this section, we first describe the three domain knowledge settings and the rationale for selecting these settings. Then, we describe the design of the user study.

\subsection{Domain knowledge settings}
\label{sec:settings}

\begin{figure}[ht!]
    \includegraphics[width=\columnwidth]{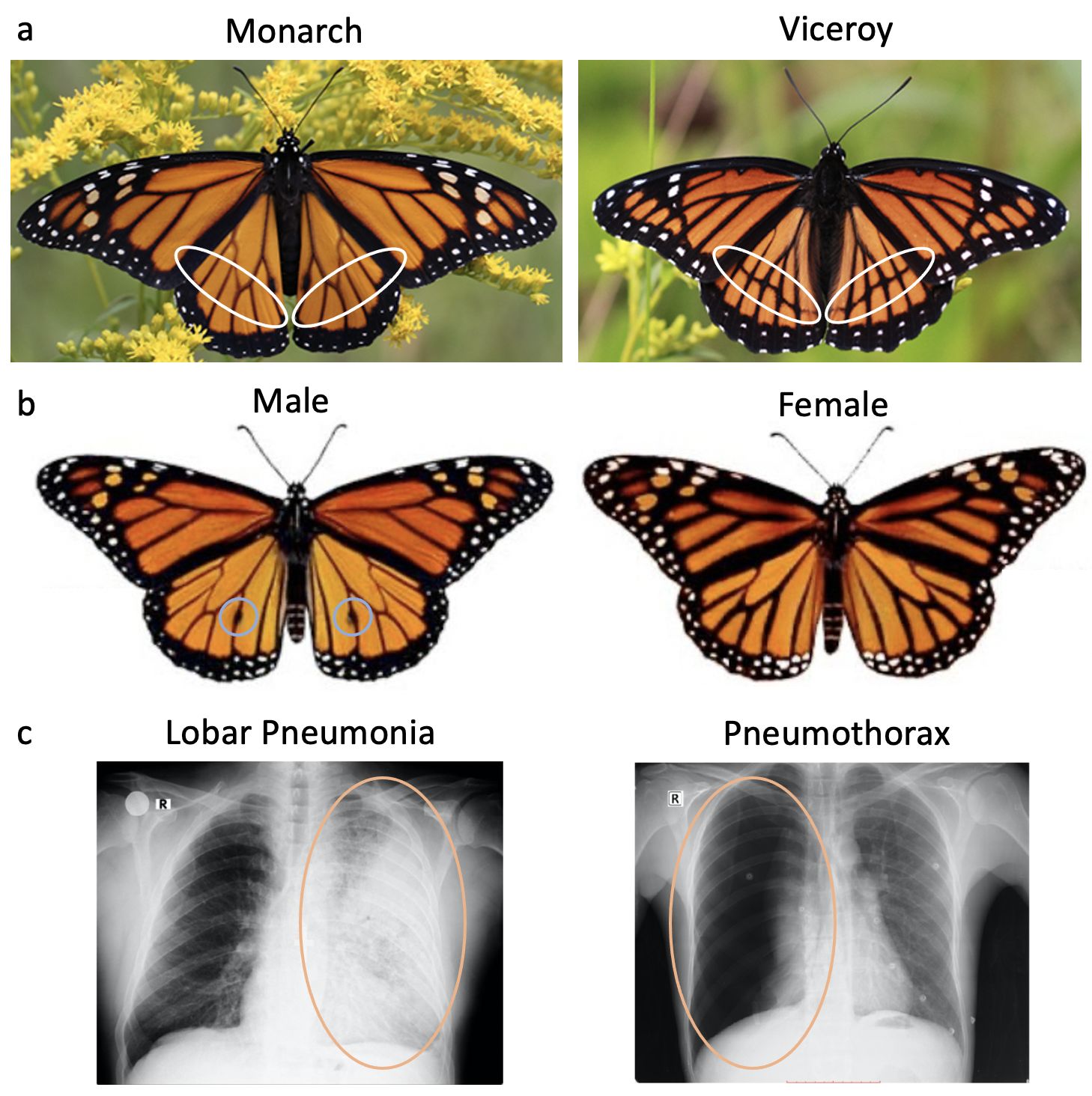}
    \vspace{-0.5cm}
    \caption{The three domain knowledge settings of the user study. (a) Monarch vs. viceroy butterflies. Viceroy butterflies have horizontal lines on the hindwings while monarch butterflies do not. (b) Male vs. female monarch butterflies. Male butterflies have black spots on the hindwings and thin veins while female butterflies do not have black spots and have thick veins. (c) Lobar pneumonia vs. pneumothorax in chest x-rays. Lobar pneumonia presents as white opacity in a lobe of the lung while pneumothorax presents as a darker region of the lung.}
    \label{fig:three-settings}
    \vspace{-0.2cm}
\end{figure}

The study spans three domain knowledge settings each involving a binary classification task: distinguishing between 1) male and female monarch butterflies, 2) monarch and viceroy butterflies, and 3) lobar pneumonia and pneumothorax in chest x-rays. For each setting, examples of each class and the distinction between the two classes are described in Figure~\ref{fig:three-settings}. All three settings are tasks that require domain knowledge to perform.

The first two settings (butterfly sexes and species) are selected because they are tasks that require domain knowledge that most people do not have, but that can be easily taught. We leverage this aspect of these two settings to control the laypeople and domain expert groups by teaching or not teaching a participant group. Specifically, in the first setting (butterfly sexes), we teach the distinction to students, but not to doctors, making students domain experts and doctors laypeople. This setting serves as a positive control: the domain expert and laypeople groups are engineered and differences (if any) between the two groups would be expected to arise in this setting. In the second setting, we teach the distinction to both students and doctors, making both groups domain experts. This setting serves as a negative control: both groups are engineered to be domain experts and no difference between the two groups is expected.

The third setting (chest x-rays) is selected because it is a task that requires medical domain knowledge. It is a relatively simple medical task that doctors, i.e., people who have a medical degree, are expected to know but that others would not be expected to know. In this setting, we do not provide any teaching. Doctors, having learned the domain knowledge in medical school, are domain experts, while students are laypeople. This setting is the experimental condition: the domain expert and laypeople groups arise from natural differences and differences (if any) between the two groups are expected to resemble those of the positive control.

The logic and expected results of the three settings described above are summarized in Table~\ref{table:study-design}.

\subsection{User study design}

\begin{table*}[ht!]
\centering
\begin{tabular}{l|l|l|l}
Setting                                                           & Task                                                                         & Groups                                                                                                & Expected result                                                                                         \\
\midrule
\begin{tabular}[c]{@{}l@{}}Positive\\ control\end{tabular}        & \begin{tabular}[c]{@{}l@{}}Male vs. female \\ monarch butterfly\end{tabular} & \begin{tabular}[c]{@{}l@{}}DE: students \\ (with teaching)\\ LP: doctors\end{tabular}                 & Differences between groups                                                                              \\
\midrule
\begin{tabular}[c]{@{}l@{}}Negative\\ control\end{tabular}        & \begin{tabular}[c]{@{}l@{}}Monarch vs. viceroy \\ butterfly\end{tabular}     & \begin{tabular}[c]{@{}l@{}}DE: students and doctors \\ (both groups \\ receive teaching)\end{tabular} & No differences between groups                                                                           \\
\midrule
\begin{tabular}[c]{@{}l@{}}Experimental \\ condition\end{tabular} & \begin{tabular}[c]{@{}l@{}}Pneumothorax vs. \\ lobar pneumonia\end{tabular}  & \begin{tabular}[c]{@{}l@{}}DE: doctors\\ LP: students\end{tabular}                                    & \begin{tabular}[c]{@{}l@{}}Similar to positive control; \\ different from negative control\end{tabular}
\end{tabular}
\caption{Study design. This table describes the three settings in the study, their tasks, domain experts (DE) and laypeople (LP) groups, and expected results.}
\label{table:study-design}
\end{table*}

\begin{figure*}[ht!]
    \includegraphics[width=2\columnwidth]{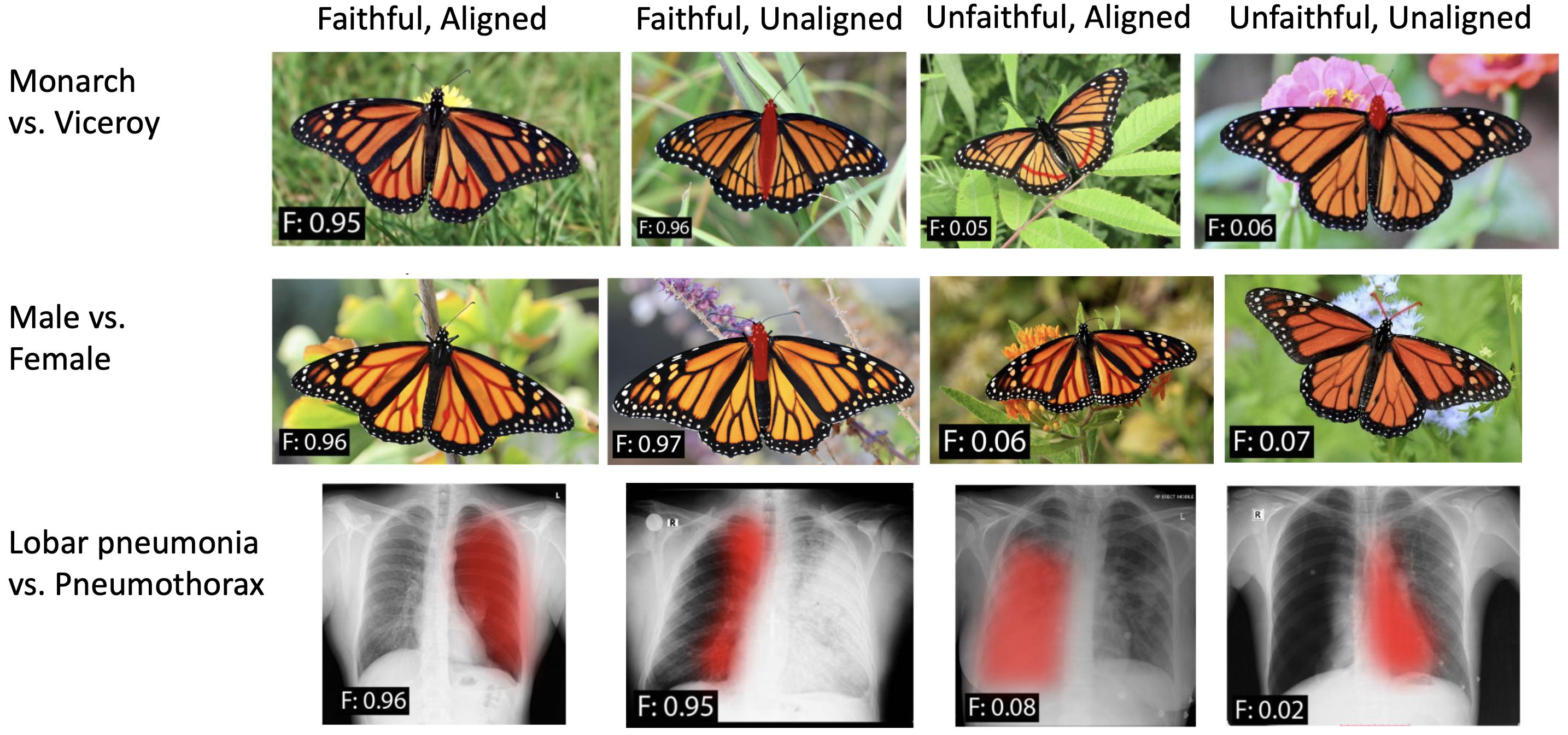}
    \caption{Explanations for each setting. Explanations can be faithful or unfaithful (i.e., accurately or inaccurately capturing the behavior of the underlying model), and aligned or unaligned (i.e., consistent or inconsistent with prior knowledge). Each setting has four types of explanations: faithful and aligned, faithful and unaligned, unfaithful and aligned, and unfaithful and unaligned. }
    \label{fig:explanations}
\end{figure*}

\begin{table*}[ht!]
\centering
\begin{tabular}{l|l|l|l}
Example             & Trust score & Expected trust for LP & Expected trust for DE \\
\midrule
Faithful, aligned   & A           & Neutral or recommend         & Recommend                        \\
Faithful, unaligned & B           & Neutral or recommend         & Reject                           \\
Unfaithful, aligned & C           & Neutral or reject            & Recommend                        \\
Unfaithful, unaligned & D           & Neutral or reject            & Reject                          
\end{tabular}
\caption{Analysis logic. This table outlines the four different scenarios in each of the three settings and the expected trust for laypeople (LP) and domain experts (DE). In the table, "recommend" refers to both "mildy recommend" and "recommend" the model, and "reject" refers to both "midly reject" and "reject" the model.}
\label{table:analysis-logic}
\end{table*}

\setlength{\parskip}{0pt}

In the study, participants are first provided background information on convolutional neural networks performing binary classification and post hoc explanations, including how to interpret the F-score and salient features of an explanation. Then, participants are presented with three binary classification settings: distinguishing between monarch and viceroy butterflies, between male and female monarch butterflies, and between pneumonia and pneumothorax in chest x-rays. As discussed in Section~\ref{sec:settings}, the first two settings serve as two simple tasks that are not commonly known but that are easy to learn and the third setting serves as a domain-specific task in which doctors are domain experts. The setup and purpose of each setting are described in Table~\ref{table:study-design}.

For each setting, participants are asked to imagine that they are hired by a company or regulatory agency to help the organization decide whether to recommend or reject models. Depending on the setting and participant group, participants sometimes received teaching to become domain experts in the setting, as described in Table~\ref{table:study-design}. Each setting has four images. Participants are first shown the image and asked to make a classification on their own. Then, participants are shown an explanation for the model's prediction of the image class and asked whether they would recommend or reject the model, a question that measures their trust in the model. An explanation consists of an F-score and salient regions of the image and is either strongly or weakly faithful and aligned or unaligned (resulting in four combinations). Images of the four scenarios for each setting are shown in Figure~\ref{fig:explanations}. When making a recommendation, participants are asked to select among the following five options: recommend the model, mildly recommend the model, neutral, mildly reject the model, and reject the model. To avoid assessments of model trust carrying over from one example to the next, participants are told that each of the four images involves a different model.

The models and explanations in the user study are hypothetical and engineered to allow for precise control. Participants were told that the accuracy for all models is 97\%, saliency maps were created using Adobe Illustrator to highlight clearly aligned or unaligned regions of an image, and explanations were assigned either extremely low (e.g., 0.05) or extremely high (e.g., 0.95) F-scores. 

After the three settings, participants were asked a three open-ended questions about their thought process when determining whether to recommend or reject a model, additional information they would find useful when determining the extent to which to trust a model, and whether they prefer more interpretable (less accurate) models or more accurate (less interpretable) models in medicine.

We recruited doctors by emailing doctors listed on public hospital webpages (all doctors at Massachusetts Eye and Ear Infirmary and all radiologists at Massachusetts General Hospital). We recruited students by emailing professors of Spring 2023 computer science courses at our university and asking them to disseminate the survey to their students. 

%% file: 5-exp-results.tex
\begin{figure*}[ht!]
    \includegraphics[width=2\columnwidth]{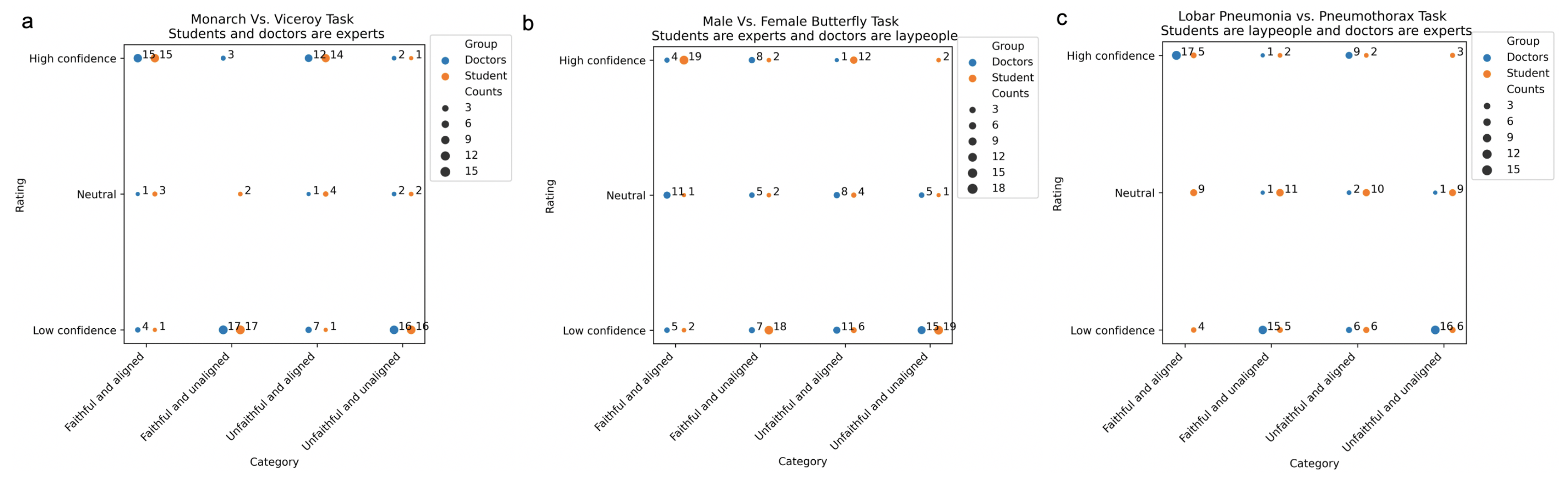}
    \caption{Distribution of responses among laypeople and domain expert groups in the (a) negative control, (b) positive control, and (c) experimental conditions. The results indicate that for laypeople, trust is associated with explanation faithfulness, while for domain experts, trust is associated with explanation alignment. The positive control, which engineers group differences, and the experimental condition, which naturally has these group differences, yield the same results.}
    \label{fig:results}
\end{figure*}

\section{Analyses and Results}

We obtained responses from 22 students and 22 doctors. In order to measure trust quantitatively, we convert the qualitative responses to numbers (``trust score''): recommend the model (2), mildly recommend the model (1), neutral (0), mildly reject the model (-1), and reject the model (-2).

We perform analyses of individual differences, group differences, and open-ended responses. Each set of analyses is described in the subsections below. Table~\ref{table:analysis-logic} presents information and logic that inform our analyses, and we will refer to this content when explaining the analyses.

\subsection{Individual differences}
For laypeople, we hypothesize that trust is based on explanation faithfulness because, by definition, laypeople do not have relevant domain knowledge and thus do not understand explanation alignment. Thus, in Table~\ref{table:analysis-logic}, we expect differences A-B and C-D to be zero on average (because faithfulness is constant) and differences A-C and B-D to be positive on average (because higher faithfulness is hypothesized to lead to higher trust). For each setting, we filter laypeople responses, keeping only responses from participants who classified all four examples incorrectly (i.e., participants for whom the laypeople assumption is valid). Group sizes after filtering are in the Appendix (Table~\ref{table:counts-filtering}). Then, for the two settings involving laypeople, we calculate these differences for each layperson ("individual differences"), and test the null hypothesis that the mean of the distribution is zero using a two-sided one-sample t-test. 

For domain experts, we hypothesize that trust is based on explanation alignment because domain experts, due to their strong prior knowledge, may experience a bias that causes them to focus more on whether the explanation aligns with what they personally know and less on the faithfulness of the explanation with respect to the model. Thus, in Table~\ref{table:analysis-logic}, we expect differences A-C and B-D to be zero on average (because alignment is constant) and differences A-B and C-D to be positive on average (because higher alignment is hypothesized to lead to higher trust). For each setting, we filter domain expert responses, keeping only responses from participants who classified all four examples correctly on their own (i.e., participants for whom the domain expert assumption is valid). Group sizes after filtering are in the Appendix (Table~\ref{table:counts-filtering}). Then, for each setting, we calculate these differences for each domain expert ("individual differences"), and test the null hypothesis that the mean of the distribution is zero using a two-sided one-sample t-test. 

Throughout this study, we performed a total of 36 hypothesis tests. To adjust for multiple testing, we use the Bonferroni correction. Therefore, conclusions for all hypothesis tests are based on a significance threshold of 0.0001 (= 0.05/36).

The distribution of responses in each setting is shown in Figure~\ref{fig:results}. The results of the hypothesis tests are shown in Table~\ref{table:pvals-indiv-diff}. We find that for laypeople, trust tends to be associated with explanation faithfulness while for domain experts, trust tends to be associated with explanation alignment. The results for laypeople are not all statistically significant, but all the results for domain experts are statistically significant. Note that results of the experimental condition (pneumonia vs. pneumothorax setting), where the laypeople and domain expert groups are natural, reproduce results from the positive control (male vs. female monarch butterfly setting), where the laypeople and domain expert groups are engineered.
All together, these results are consistent with our hypotheses.

\begin{table*}[ht!]
\centerline{
\begin{tabular}{| *{9}{l|} }
    \hline
    Comparisons   
    & \multicolumn{2}{c|}{Positive control (LP / DE)}
    & \multicolumn{2}{c|}{Negative control (DE / DE)}
    & \multicolumn{2}{c|}{Experimental condition (LP / DE)}   \\
    \hline
    A-B   
    &   -0.1 
    &   1.5*  
    &   1.63*  
    &   1.25*  
    &   0.23  
    &   1.81*   \\
    \hline
    C-D   
    &  0.25  
    &  1.05*  
    &  1.47*
    &  0.95*  
    &  0
    &  1.11*   \\
    \hline
    A-C 
    & 0.45 
    & 0.5  
    & 0.05     
    & 0.3      
    & 0.32     
    & 0.82              \\
    \hline
    B-D   
    & 0.8*   
    & 0.05    
    & -0.11     
    & 0    
    & 0.09      
    & 0.11            \\
    \hline
\end{tabular}}
\caption{Individual differences in post hoc explanation trust. For each setting, the table shows individual differences in trust scores for laypeople (LP) and  domain experts (DE). *indicates a statistically significant result at a significance threshold of 0.0001 (adjusted for multiple testing). Exact t-statistics and p-values can be found in the Appendix. A vertical bar separates results for the two participant groups in each setting. Results indicate that laypeople tend to base their trust on explanation faithfulness (although results are not always statistically significant) while domain experts tend to base their trust on explanation alignment (results always statistically significant).}
\label{table:trust-scores-indiv-diff}
\end{table*}

\subsection{Group differences}

\begin{table*}[h!]
\centerline{
\begin{tabular}{l|l|l|l}
    & Positive control & Negative control & Experimental condition \\
    \midrule
    A-A & -0.95*       & -0.13        & 0.90*              \\
    \midrule
    B-B & 0.65        & 0.47        & -0.81              \\
    \midrule
    C-C & -0.65        & -0.33        & 0.91              \\
    \midrule
    D-D & 0.25        & 0.33        & -0.55*       
\end{tabular}}
\caption{Group differences in post hoc explanation trust between the two participant groups in each setting. For each setting, the table shows group differences in trust scores. *indicates a statistically significant result at a significance threshold of 0.0001 (adjusted for multiple testing). Exact t-statistics and p-values can be found in the Appendix.}
\label{table:trust-scores-group-diff}
\end{table*}

For laypeople and domain experts, we also hypothesize specific levels of trust, depending on explanation faithfulness and alignment, as described in Table~\ref{table:analysis-logic}. Based on these hypotheses, we expect the difference between the average domain expert response and average layperson response to be positive for when the explanation is faithful/aligned or unfaithful/aligned, and we expect this difference to be negative when the explanation is faithful/unaligned or unfaithful/unaligned. For the negative control, we expect no difference between the two groups (both domain experts). For each setting, we calculate the average trust score for one group (domain experts) and the average trust score for the other group (laypeople in the positive control and experimental condition, and domain experts in the negative control), and test the null hypothesis that the two population means are equal using a two-sided two-sample t-test.  

The results are shown in Table~\ref{table:trust-scores-group-diff}. As expected, for the negative control, there were no significant differences. For the positive control, differences are statistically significant for the faithful/aligned scenario. For the experimental condition, differences are statistically significant for the faithful/aligned and unfaithful/unaligned scenarios. While many tests did have relatively low p-values, these p-values were not statistically significant at the stringent confidence level that was adjusted for multiple testing.


\subsection{Open-ended responses}

Lastly, we analyzed the participants' responses to the open-ended questions. When asked to explain their thought process, students (laypeople in the medical setting) tended to focus on explanation faithfulness while doctors (domain experts in the medical setting) tended to focus on explanation alignment. Example student responses include ``for the lungs, I used the F score because I didn't know whether what it highlighted was correct'', "if the F scores for an area aligned positively with the human-centric explaination", and ``I looked for models whose explanations matched the background knowledge I was given and had high F-scores''. Example doctor responses include ``When I know what exists (ptx and pn) and I see the heat maps. That too [sic] me is more important than any F value stats'', ``Whether the heat map corresponded with the relevant area to make the appropriate classification'', ``whether the model looked at the relevant anatomy''. Thus, participants' qualitative descriptions of their thought process are consistent with the quantitative findings discussed above.

When asked whether they prefer (A) models that are more accurate but less interpretable, (B) models that are more interpretable but less accurate in medicine, or (C) ``It depends'', students tended to choose all three answer options with similar frequency while doctors tended to prefer option C. Out of the 22 students, 7 responded A, 8 responded B, and 7 responded ``It depends''. Out of the 21 doctors, 6 responded A, 3 responded B, and 12 responded ``It depends''. These results are shown in the Appendix (Table~\ref{table:interpretable_result}). A chi-square test for association indicated that the association was not statistically significant (chi-square = 3.64, p-value = 0.16). Doctors who answered ``It depends'' emphasized in their responses that some clinical scenarios require accuracy more than interpretability especially when domain knowledge is limited. They also emphasized the need for a representation of the confidence of a model in its prediction when making a decision. Students who answered "It depends" also emphasized these points, adding ethical concerns also need to be taken into account as some problems may have more relevant ethical concerns than others.

%% file: 6-conclusion.tex
\section{Conclusion}

In summary, to investigate how the faithfulness and alignment of post hoc explanations affect model trust in laypeople and domain experts, we conducted a user study with computer science students and doctors. We find that for laypeople, trust is associated with explanation faithfulness, while for domain experts trust is associated with explanation alignment. We hypothesize that this is because domain experts, with their strong domain knowledge may experience a cognitive bias, causing them to focus on the alignment of explanations with prior knowledge, while laypeople, with no prior knowledge, have only explanation faithfulness to base their trust on.

Explanation faithfulness is a fundamental and foremost criterion of a model explanation: a user ought to first establish that an explanation is faithful, i.e., accurately represents the underlying model, before interpreting information in the explanation, such as explanation alignment. In this study, we find that laypeople correctly focus on explanation faithfulness, while domain experts tend to focus on explanation alignment without first checking explanation faithfulness.

One potential limitation of this study is that it uses engineered post hoc explanations to precisely design explanations that are faithful or unfaithful and aligned or unaligned. While these engineered explanations enable us to draw precise conclusions, they may not generalize perfectly to real-world post hoc explanations. In addition, while the study includes a satisfactory number of participants, it can still benefit from a larger number of participants.

To our knowledge, this work is the first to show that (1) different aspects of post hoc explanations affect laypeople and domain experts' trust in a model and (2) domain experts are subject to specific biases due to their expertise when interpreting post hoc explanations. By uncovering this phenomenon and exposing this cognitive bias, this work motivates the need to educate end users about how to properly interpret explanations and overcome their own cognitive biases, and motivates the development of simple and interpretable faithfulness metrics for end users. This research is particularly important and timely as post hoc explanations are increasingly being used in high-stakes, real-world settings such as medicine.



%% file: 7-appendix.tex
\section*{Appendix}

\begin{table*}[ht!]
\centerline{
\begin{tabular}{| *{9}{c|} }
    \hline
    Comparisons    
    & \multicolumn{2}{c|}{Positive control (LP / DE)}
    & \multicolumn{2}{c|}{Negative control (DE / DE)}
    & \multicolumn{2}{c|}{Experimental condition (LP / DE)}   \\
    \hline
    A-B   
    &   -0.62 (0.54)  
    &   *8.18 ($5.65e^{-8}$)  
    &   *10.40 ($4.88e^{-9}$)  
    &   *5.78 ($1.43e^{-5}$)  
    &   1.23 (0.23)  
    &   *5.78 ($1.42e^{-5}$)   \\
    \hline
    C-D   
    &   1.75 (0.096)  
    & *4.32 ($2.9e^{-4}$)  
    & *7.64 ($4.72e^{-7}$)  
    & *4.50 ($2.4e^{-4}$)  
    &  0 (1.0)  
    &   *4.5 ($2.45e^{-4}$ )   \\
    \hline
    A-C 
    & 2.93 (0.009) 
    & 2.32 (0.03)  
    &  0.27 (0.79)     
    & 1.03 (0.32)      
    & 1.50 (0.15)      
    &  3.57 (0.0026)              \\
    \hline
    B-D   
    & *4.0 ($7.7e^{-4}$)      
    & 0.33 (0.75)      
    & -0.81 (0.43)      
    & 0 (1.0)       
    & 0.42 (0.68)      
    & 0.81 (0.43)            \\
    \hline
\end{tabular}}
\caption{Results of hypothesis tests for individual differences for laypeople (LP) and domain experts (DE). For each test, the table shows the t-statistic with the p-value in parentheses. *~indicates a statistically significant result at a significance threshold of 0.0001. Results indicate that laypeople tend to base their trust on explanation faithfulness (although results are not always statistically significant) while domain experts tend to base their trust on explanation alignment (results always statistically significant)}
\label{table:pvals-indiv-diff}
\end{table*}

\begin{table}[h!]
\centerline{
\begin{tabular}{l|l|l|l}
    & Positive control & Negative control & Experimental condition \\
\midrule
A-A & *-4.1 ($1.9e^{-4})$        & -0.82 (0.41)        & *5.4 ($6.3e^{-6}$)              \\
\midrule
B-B & 3.3 (0.002)        & 1.07 (0.29)        & -3.4($1.9e^{-3}$)              \\
\midrule
C-C & 3.27 (0.002)        & -1.7 (0.1)        & 1.5(0.15)              \\
\midrule
D-D & 0.14 (0.89)        & 0.47 (0.64)        & *-4.3($1.5e^{-4})$       
\end{tabular}}
\caption{Results of hypothesis tests for group differences. For each test, the table shows the t-statistic with the p-value in parentheses. *indicates a statistically significant result at a significance threshold of 0.0001 (adjusted for multiple testing).}
\label{table:pvals-group-diff}
\end{table}

\begin{table}[h!]
\centerline{
\begin{tabular}{l|l|l|l}
    Setting & Students & Doctors \\
\midrule
Positive control & 22 $\rightarrow$ 22      & 22 $\rightarrow$ 20                    \\
\midrule
Negative control & 22 $\rightarrow$ 19        & 22 $\rightarrow$ 20                      \\
\midrule
Experimental condition & 22 $\rightarrow$ 18        & 22 $\rightarrow$ 17                  \\
\midrule
    
\end{tabular}}
\caption{Number of participants in domain expert group before and after filtering. (Notation is: Before number $\rightarrow$ After number)}
\label{table:counts-filtering}
\end{table}

\begin{table}[h!]
\centerline{
\begin{tabular}{l|l|l|l}
    & Prefer accurate models & Prefer intepretable models & It depends \\
\midrule
Students & 7       & 8        & 7             \\
\midrule
Doctors & 6        & 3        & 12              \\
\midrule      
\end{tabular}}
\caption{Counts for number of responses for students versus doctors for interpretability question. Doctors tend to say "It depends" more than students where it is more uniform.}
\label{table:interpretable_result}
\end{table}